\documentclass[a4paper]{mem}
\usepackage{natbib}
\usepackage{graphicx}
\usepackage{txfonts}
\usepackage{balance}
\usepackage[a4paper]{hyperref}
\idline{73}{23}
\begin{document}
   \title{On the binarity of carbon-enhanced, metal-poor stars
}

   \author{S. Tsangarides \inst{1},
   S. G. Ryan \inst{1},
          \and
          T. C. Beers \inst{2}
}


   \institute{Department of Physics and Astronomy, The Open University,
Walton Hall, Milton Keynes, MK7 6AA, United Kingdom.
\email{s.tsangarides@open.ac.uk, s.g.ryan@open.ac.uk}
         \and
             Department of Physics and Astronomy and JINA: Joint
Institute for Nuclear Astrophysics, Michigan State University, East
Lansing, Michigan 48824, USA.
\email{beers@pa.msu.edu}
             }

   \abstract{
We report on a programme to monitor the radial velocities of a
sample of candidate and confirmed carbon-enhanced, metal-poor
(CEMP) stars. We observed 45 targets using the Echelle
Spectrographs of three 4-m class telescopes. Radial velocities for
these objects were calculated by cross-correlation of their
spectra with the spectrum of HD~140283, and have errors $<$ 1 km
s$^{-1}$. Sixteen of our programme's targets have reported carbon
excess, and nine of these objects also exhibit s-process
enhancements (CEMP-s). We combine these stars' radial velocities
with other literature studies in search of binarity. The search
reveals that four of our CEMP-s stars ($\sim$ 44\%) are in binary
systems. Using the analysis of Lucatello et al. (2004), we find
that all the CEMP-s stars in our sample are binaries. This
conclusion implies that CEMP-s stars may be the very metal-poor
relatives of CH and Ba II stars, which are believed to have
acquired their peculiar abundance patterns by mass transfer from a
thermally-pulsing AGB companion. \keywords{binaries: spectroscopic
-- binaries: mass transfer -- stars: Population II -- stars:
carbon} }
\authorrunning{S. Tsangarides et al.}
\titlerunning{Binarity in CEMP stars}
\maketitle
%

\section{Introduction}

Large-scale surveys of metal-poor stars, such as the HK and
Hamburg-ESO surveys, find that a large portion of their
metal-poor\footnote{We will use the term `metal-poor' to refer to
objects with [Fe/H] $^<_{\sim}$ -2.0. We adopt
[A/B]~=~log(N$_A$/N$_B$)$_{star}$~$-$~log(N$_A$/N$_B$)$_{\odot}$,
for elements A and B.} sample exhibit considerable excess of
carbon relative to iron. In fact, the numbers of these C-enhanced,
metal-poor (CEMP) stars rise as [Fe/H] decreases \citep{RBS99};
CEMP stars may account for up to $\sim$ 25\% of stars with [Fe/H]
$^<_{\sim}$ -2.0.

The enhancement of C in these stars is accompanied by at least
five different abundance patterns. \citet{ARTNBA03} report that
$\sim$ 70\% of CEMP stars in their sample exhibit super-solar
abundances of the s-process elements relative to iron (CEMP-s
stars). In addition, some objects may possess excesses in both the
s- and r-process elements (CEMP-r/s; e.g. \citealt{H+00}) and one
star (CS~22892-052) has an r-process overabundance (CEMP-r).
Finally, some objects show no excess of neutron-capture elements
(CEMP-no), but a few are rich in the $\alpha$-elements
(CEMP-$\alpha$).

\begin{table*}
\begin{flushleft}
\caption{A brief summary of our programme's six observing runs.}
\begin{tabular}{lccl}
\hline
Telescope/Instrument& AAT/UCLES& WHT/UES& TNG/SARG\\
\hline
Dates& 2000 Sep& 2001 Aug, 2002 Apr, Jun, Jul& 2002 May\\
N$_{objects}$& 3& 61& 3\\
$\lambda\lambda$ (\AA)& 3750 -- 4900& 3550 -- 5860& 3900 -- 5140\\
Resolution& 40,000& 52,000& 57,000\\
Dispersion (km s$^{-1}$ pix$^{-1}$)& 3.13& 1.75& 2.20\\
\hline
\end{tabular}
\end{flushleft}
\end{table*}

As CEMP stars are not luminous enough to be AGB objects, it is frequently
speculated that they acquired their peculiar abundance patterns by binary
mass transfer from an AGB star. In particular, CEMP-s stars bear a
surface abundance pattern similar to CH and Ba II stars (\citealt{K42};
\citealt{W65}; see also the review by O. Pols in these Proceedings).
\citet{MW90} show that all CH and Ba II stars are in
binary systems with faint, degenerate companions. \citet{JVMU98}
find that their periods vary from 100 days to 10 yr and their
eccentricities are mostly small. These orbital properties suggest
that CH and Ba II stars are produced by mass transfer from a former
thermally-pulsing AGB companion which has subsequently evolved into a
white dwarf (e.g. \citealt{HEPT95}).

Are CEMP-s stars the metal-poor relatives of CH and Ba II stars?
Although several CEMP-s stars are members of binaries, we do not
yet possess the dynamical evidence to confirm this link. Also, we
cannot yet verify that the carbon excess of the other four
CEMP-star abundance patterns is produced by the same mass-transfer
paradigm. This work will attempt to answer the simpler of the two
questions: how many CEMP-s stars are binaries? We review our
observations in $\S$2, describe the calculation of radial
velocities in $\S$3, and present our search for binaries in $\S$4.
Finally, in $\S$5, we focus our discussion to the confirmed CEMP-s
stars, and ascertain the binary fraction among them.
%

\section {Observations}

Details of the sample selection, observations and data reduction
are discussed in \citet{THESIS}. For the purposes of this work,
suffice it to say that we have observed 31 candidate CEMP stars
from the HK survey (\citealt{BPS92}, and subsequent unpublished
work), and 14 stars with prior observations, as part of a
programme to increase the sample of known CEMP stars and monitor
their radial velocities. Most candidate stars had no prior
observations at the time our programme started; thus, we had no
knowledge of their kinematics when constructing our target list.
In addition, we selected pre-observed stars only based on our
runs' right-ascension and declination windows. Thus, we expect not
to have biased the sample towards or against the inclusion of
binaries.

Our targets were observed using the Echelle Spectrographs of three
4-m class telescopes during six runs. Table 1 briefly summarises
these runs. We obtained more than 77 spectra, observing most stars
in two or more different runs. We had signal-to-noise $^>_{\sim}$
10 in the spectra used for the calculation of radial velocities,
and $\sim$ 40 for those objects whose abundances were additionally
to be calculated. Our wavelength coverage contains the G band of
CH around 4320 \AA, the CN band near 3883 \AA\ and features from
several n-capture elements.

\section{Calculation of heliocentric radial velocities}

The heliocentric radial velocities (HRVs) of our programme targets
were calculated by cross-correlation of their reduced,
sky-subtracted spectra with the spectrum of HD~140283. This star
was observed in all our runs, and was used to set the zero-point
of our velocities. Instead of adopting a literature velocity for
it, we compared the observed and rest wavelengths of 200-400 clean
metallic lines (typically Fe~I and Ti~II) to calculate its
geocentric velocity.

From six runs, our mean velocity for HD~140283 is -170.98 $\pm$
0.22 km s$^{-1}$. Given this standard deviation, we adopt a
systematic error of 0.30 km s$^{-1}$ in our HRVs, in order to
compare them with literature velocities at different zero-points
($\S$4). Our mean HRV is identical to the velocity calculated by
\citet{L+02} for this object (-170.981 $\pm$ 0.29 km s$^{-1}$),
from 19 observations spanning 8.5 yr.

Our HRV calculations are also susceptible to two internal errors.
These include the deviation of individual pairs of lines from the
mean geocentric velocity of HD~140283, and the deviation of
individual pairs of echelle orders from the mean cross-correlation
velocity of the target.

The quadrature-sum 1$\sigma$ error of internal and systematic
sources is taken to be the total error in each HRV. Total errors
are always considerably less than 1 km s$^{-1}$.

\section{The search for binaries}

We combined our HRVs with literature velocities which also had
errors $<$ 1 km s$^{-1}$, and used them to search for HRV
variation. The combined velocities were subjected to two tests: a
$\chi^2$-test (e.g. \citealt{CLSLM03}) and plots of velocity
against Julian date of observation. In total, we tested 41
objects, excluding two stars with only a single HRV and two
objects which are potential white dwarfs contaminating our sample
of candidate CEMP stars.

In the $\chi^2$-test, we assumed the mean of all HRVs weighed by
their total error to be the target's true velocity
($\langle$HRV$\rangle$). Thus, we calculated the
$\chi^2$-statistic using:
\begin{displaymath}
\chi^2 = \sum_{i=1}^{n} \frac{HRV_i - \langle HRV \rangle}{\sigma_i}  .\\
\end{displaymath}
In addition, we adopted the degrees of freedom ($f$) to be the
number of observations less 1, and used the $\chi^2$ distribution
to evaluate the probability ($p$) of observing the
$\chi^2$-statistic given $f$. Clearly, $p$ is equivalent to the
probability that individual HRVs are all measures of the object's
true velocity within the observational errors.

The $\chi^2$-test becomes sensitive to even minor HRV variations
for $f \leq$ 4. That is, even small differences between velocities
at this $f$, cause the $\chi^2$ statistic to become
unrealistically large, underestimating the $p$-value and biasing
the test towards the detection of possibly non-real binarity. As
most of our targets have only a few observations, however, they
lie in this range of $f$. Thus, for the $\chi^2$ test, we consider
that a target exhibits HRV variation if $p <$ 1\% at $f >$ 4, and
is only potentially variable for $p ^<_{\sim}$ 10\% if $f \leq$ 4.

Further to the $\chi^2$-test, we plotted each star's HRVs against
their Julian date of observation (HJD). Since HRV variation does
not necessarily imply binarity, we classified an object as binary
only if the amplitude of HRV variation was large enough to be
discerned visually from these plots.

Figure 1 illustrates the four classes of HRV variability we were
able to identify. The filled triangles are literature HRVs (Table
2) and the open circles are velocities measured in $\S$3. The
dashed line through the data is \textit{not} a systemic velocity;
even for the confirmed binaries, we usually lack enough
observations to calculate a unique orbital solution. Instead, the
line represents the mean HRV and is only meant to aid the eye in
detecting velocity variation.

   \begin{figure*}
   \centering
\resizebox{\hsize}{!}{\rotatebox[]{0}{\includegraphics{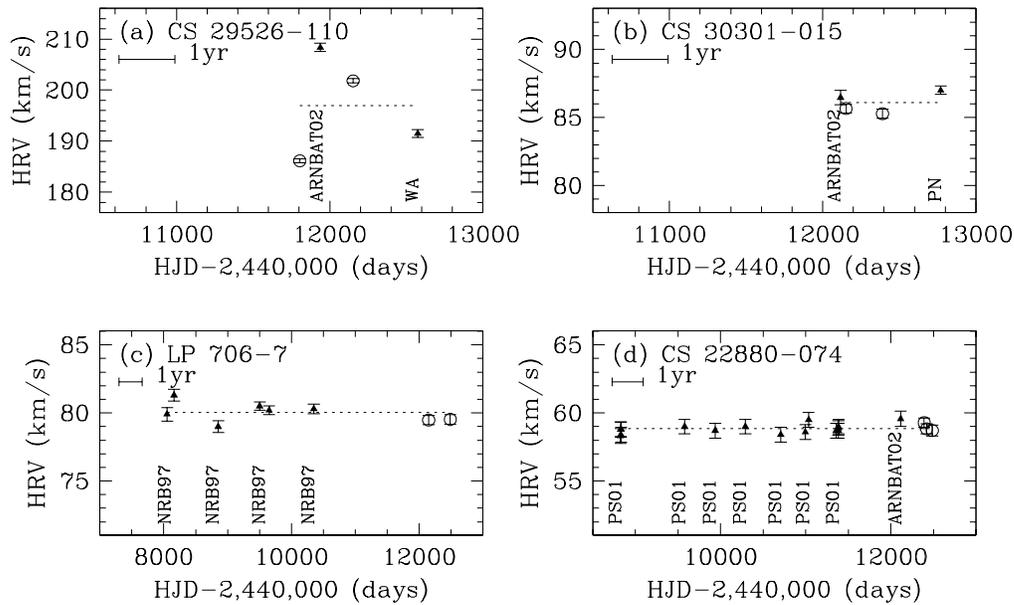}}}
   \caption{Plots of heliocentric radial velocity (HRV) in km s$^{-1}$
versus Julian date of observation (HJD) in days, for 4 stars in
our sample. The line through the data is simply the arithmetic
mean HRV of all the observations. The filled triangles represent
literature velocities and the open circles are velocities
calculated in $\S$3. Error bars are quadrature-sum errors of each
velocity's internal errors and our adopted systematic error.}
     \label{FigGam}
   \end{figure*}

Plotted in panel (a) is CS~29526-110, an object reported to be
CEMP-s (Table 2). Even with $f$ = 3, this object clearly is binary
since the amplitude of HRV variation is $\sim$ 10 km s$^{-1}$, or
$>$ 25$\sigma$ away from the mean line. Of the 41 tested objects,
we found 13 binaries, all as visibly clear as CS~29526-110. Note
that these objects exhibit $p$ = 7 $\times$ 10$^{-6}$ to
10$^{-210}$.

In panel (b) we show the potential HRV-variable CEMP-s star
CS~30301-015. This object has $p$ = 0.001, but with $f$ = 3, the
variation is still not large enough (within 3$\sigma$) to confirm
variability. This object requires further HRV monitoring. In
total, 10 objects were classified as potential variables.

The seemingly HRV-variable, CEMP-s star LP~706-7 is shown in panel
(c). HRV variation is detected in the $\chi^2$-test ($p$ = 0.002;
$f$ = 7), but its amplitude is not large enough to classify this
object as binary from inspection of the plot, since the reported
HRVs lie within 3$\sigma$ of the mean
velocity\footnote{\textit{If} the observed variation is real, its
small amplitude may be caused by stellar pulsation, spots or
planetary companions, instead of a stellar companion. Discussion
on these topics is deferred to \citet{THESIS}, as the present work
only focuses on the large-amplitude, confirmed binaries.}. More
likely, an underestimation of our errors could have resulted in a
larger $\chi^2$-statistic, thereby leading the $\chi^2$ test to
detect non-existent variation. We identified 3 such objects, the
other two being BS~16090-048 and CS~22892-052. The latter was also
studied by \citet{PS01}, who comment that any periodicity in this
object's HRV variation requires confirmation.

Finally, panel (d) shows the CEMP-s star CS~22880-074, which
exhibits no HRV variation over a period of $\sim$ 4000 days. This
object is either single or, if in a binary, the orbital axis of
the system is inclined at 0$^o$. 15 stars exhibit no apparent
variation, but most have only a few observations.

\section{Discussion}

If we assume all 41 tested objects to be CEMP stars, we find a
binary frequency of $\sim$ 32\%. However, only 16 objects in our
sample have \textit{reported} C overabundances in the literature,
and 9 of these possess s-process enhancements. Table 2 summarises
abundances and binary status for these 16 confirmed CEMP objects.
The first column gives the star name, and the second through fifth
columns [Fe/H], [C/Fe], [Ba/Fe] and [Pb/Fe] abundances,
respectively. The sixth column shows the result of our two
binarity tests: SB1 is used to label the binaries, vbl for objects
seeming to exhibit HRV variation but not confirmed binarity, vbl?
for potential HRV variables, and non-vbl to identify objects with
no apparent HRV variation. The final column gives references from
which we obtained abundances and HRVs.

\begin{table*}
\begin{flushleft}
\caption{Abundances and binary status for 16 confirmed CEMP stars.}
\begin{tabular}{lcccccl}
\hline
Star& [Fe/H]& [C/Fe]& [Ba/Fe]& [Pb/Fe]& Binarity test& Refs\\
\hline
BS~16080-175& -1.86& +1.75& ...& ...& vbl?& 1\\
BS~16929-005& -3.09& +0.92& -0.59& ...& non-vbl& 1,10,13\\
BS~17436-058& -1.78& +1.50& ...& ...& non-vbl& 1,13\\
CS~22183-015& -2.85& +2.34& +2.09& +3.17& vbl?& 1,11\\
CS~22877-001& -2.72& +1.00& -0.49& ...& vbl?& 1,4,13\\
CS~22880-074& -1.76& +1.51& +1.34& ...& non-vbl& 1,6,14\\
CS~22887-048& -1.70& +1.84& ...& ...& SB1& 1,13\\
CS~22892-052& -2.92& +0.91& +0.92& ...& vbl& 1,8,10,11,12,13,14\\
CS~22898-027& -2.25& +2.20& +2.23& +2.84& non-vbl& 1,6,13,14\\
CS~29502-092& -2.76& +1.00& -0.82& ...& non-vbl& 1,4,7,13\\
CS~29526-110& -2.38& +2.20& +2.11& +3.30& SB1& 1,6,7\\
CS~30301-015& -2.64& +1.60& +1.45& +1.70& vbl?& 1,6,13\\
CS~31062-050& -2.32& +2.00& +2.30& +2.90& SB1& 1,6,9\\
HD~196944& -2.26& +1.20& +1.10& +1.90& SB1& 1,6,7,15\\
LP~625-44& -2.71& +2.10& +2.74& +2.55& SB1& 1,2,3,5,9,12\\
LP~706-7& -2.74& +2.15& +2.01& +2.28& vbl& 1,3,12\\
\hline
\end{tabular}
\begin{tabular}{l l}
1: This work; \citet{THESIS}.& 9: \citet{A+03}.\\
2: \citet{ANRBA00}.& 10: Honda et al. (2004a,b).\\
3: \citet{A+01}.& 11: \citet{JB02}.\\
4: \citet{ANRBA02}.& 12: \citet[~NRB97]{NRB97}.\\
5: \citet{A+02}.& 13: P. North (2003, private communication; PN).\\
6: \citet[~ARNBAT02]{ARNBAT02}.& 14: \citet[~PS01]{PS01}.\\
7: W. Aoki (2003, private communication; WA).& 15: \citet{VGJP03}.\\
8: \citet{AHBS03}.&  \\
\end{tabular}
\end{flushleft}
\end{table*}

From Table 2, 4 CEMP-s stars ($\sim$ 44\%) are members of binary
systems. Most of the s-process-rich stars reported here are a
subset of Lucatello et al.'s (2004) sample. These authors
calculate expected detection rates for binary CEMP-s stars
assuming different true binary fractions. Using their analysis, we
find that our detection rate of 44\% implies that all CEMP-s stars
in our sample are likely binaries. Although this result does not
constitute enough dynamical evidence to verify that the CEMP-s
abundance pattern is produced by mass transfer from a former AGB
star, it strengthens the relation between CEMP-s, CH and Ba II
stars. Further HRV monitoring will enable the derivation of
orbital solutions, which may help establish the evolutionary
status of the companion during mass transfer.

We are still calculating abundances for three of the remaining
seven CEMP stars in Table 2, including the only other confirmed
binary (CS~22887-048). Thus, we defer discussion on these objects
to a later publication (e.g. \citealt{THESIS}).

The final four objects include the CEMP-r star, CS~22892-052, which was
briefly discussed in $\S$4, and three stars from \citet{HAKABIST04}
with no apparent n-capture or $\alpha$-element overabundances. Although
all three CEMP-no stars exhibit no HRV variability, this sample is too
small and the considered observations too few for meaningful statistics.
Thus, it is important to continue monitoring the velocities of all CEMP
objects and find more objects of the less populous abundance patterns.
Observations for both aims are planned for the future.

\begin{acknowledgements}
The authors wish to acknowledge fruitful discussions on the
binarity of CEMP stars with W. Aoki, S. Lucatello, J. E. Norris
and P. North. ST is grateful to P. North for making radial
velocities available prior to publication. ST acknowledges partial
support from a UK Universities Overseas Research Student Award
(ORS/2001031002) to his PhD programme and a travel grant to the TNG
provided by the European Commission through the ``Improving Human
Potential Programme'' awarded to the Instituto de Astrof\'isika de
Canarias. TCB acknowledges grants AST 00-98549, AST 00-98508, and
PHY 02-16783, Physics Frontier Centers/JINA: Joint Institute for
Nuclear Astrophysics, awarded by the U.S. National Science
Foundation.
\end{acknowledgements}
\balance
\bibliographystyle{aa}

\end{document}